\documentclass[conference]{IEEEtranUDE}

\usepackage{acronym}
\usepackage{acronym.conferences.telecom}
\usepackage{graphicx}

\usepackage{times}
\usepackage{amsfonts,subfigs}
\usepackage[latin1]{inputenc}
\usepackage{multirow}
\usepackage{amsmath}

\newacronym{BER}{Bit Error Rate}
\newacronym{BS}{Base Station}
\newacronym{CIR}{Channel Impulse Response}
\newacronym{CFR}{Channel Frequency Response}
\newacronym{CP}{Cyclic-Prefix}
\newacronym{DF}{Decision-Feedback}
\newacronym{FDE}{Frequency Domain Equalization}
\newacronym{FFT}{Fast Fourier Transform}
\newacronym{IDFT}{Inverse Discrete Fourier Transform}
\newacronym{ISI}{Inter-Symbol Interference}
\newacronym{LDB}{Linear Detection Bound}
\newacronym{MF}{Matched Filter}
\newacronym{MFB}{Matched-Filter Bound}
\newacronym{ML}{Maximum Likelihood}
\newacronym{MLD}{Maximum-Likelihood Detection}
\newacronym{MLDB}{Maximum-Likelihood Detection Bound}
\newacronym{MIMO}{Multi-Input Multi-Output}
\newacronym{MRC}{Maximal Ratio Combining}
\newacronym{MU}{Multi-User}
\newacronym{MUI}{Multi-User Interference}
\newacronym{MSI}{Multi-Stream Interference}
\newacronym{MUD}{MultiUser Detection}
\newacronym{MT}{Mobile Terminal}
\newacronym{MMSE}{Minimum Mean-Squared Error}
\newacronym{OFDM}{Orthogonal Frequency Division Multiplexing}
\newacronym{QAM}{Quadrature Amplitude Modulation}
\newacronym{SINR}{Signal-to-Interference-plus-Noise-Ratio}
\newacronym{SC}{Single Carrier}
\newacronym{SU}{Single User}
\newacronym{TX}{Transmitter}
\newacronym{RX}{Receiver}
\newacronym{SIMO}{Single-Input Multi-Output}
\newacronym{ZF}{Zero Forcing}

\begin{document}
%
\title{On Detection Issues in the SC-based Uplink of a MU-MIMO System with a Large Number of BS Antennas}

\author{$^{(1)}$Paulo Torres, $^{(2)}$Luis Charrua, $^{(3)}$Antonio Gusmao\\
$^{(1,2,3)}$IST - Instituto Superior Tecnico, Universidade de Lisboa, Portugal \\
$^{(1)}$EST - Escola Superior de Tecnologia de Castelo Branco, Portugal \\
$^{(1)}$paulo.torres@ipcb.pt, $^{(2)}$luischarrua@enautica.pt, $^{(3)}$gus@ist.utl.pt}

\maketitle

\begin{abstract}

This paper deals with \SC/\FDE\ within a \MU-\MIMO\ system where a large number of \BS\ antennas is adopted. In this context, either linear or reduced-complexity iterative \DF\ detection techniques are considered. Regarding performance evaluation by simulation, appropriate semi-analytical methods are proposed.

This paper includes a detailed evaluation of \BER\ performances for uncoded 4-Quadrature Amplitude Modulation (4-QAM) schemes and a \MU-\MIMO\ channel with uncorrelated Rayleigh fading. The accuracy of performance results obtained through the semi-analytical simulation methods is assessed by means of parallel conventional Monte Carlo simulations, under the assumptions of perfect power control and perfect channel estimation. The performance results are discussed in detail, with the help of selected performance bounds. We emphasize that a moderately large number of \BS\ antennas is enough to closely approximate the \SIMO\ \MFB\ performance, especially when using the suggested low-complexity iterative \DF\ technique, which does not require matrix inversion operations. We also emphasize the achievable "massive \MIMO" effects, even for strongly reduced-complexity linear detection techniques, provided that the number of BS antennas is much higher than the number of antennas which are jointly employed in the terminals of the multiple autonomous users.

\end{abstract}


\section{Introduction}\label{sec_introd}

\CP-assisted  block transmission schemes were proposed and developed, in the last two decades, for broadband wireless systems, which have to deal with strongly frequency-selective fading channel conditions. These schemes take advantage of current low-cost, flexible, \FFT-based signal processing technology, with both \OFDM\ and \SC/\FDE\ alternative choices \cite{Sari94,GC00,GC03}. Mixed air interface solutions, with \OFDM\ for the downlink and \SC/\FDE\ for the uplink, as proposed in \cite{GC00}, are now widely accepted; the main reason for replacing \OFDM\ by \SC/\FDE, with regard to uplink transmission, is the lower envelope fluctuation of the transmitted signals when data symbols are directly defined in the time domain, leading to reduced power amplification problems at the mobile terminals.

The development of \MIMO\ technologies has been crucial for the "success story" of broadband wireless communications in the last two decades. Through spatial multiplexing schemes, following and extending ideas early presented in \cite{Foschini03}, \MIMO\ systems are currently able to provide very high bandwidth efficiencies and a reliable radiotransmission at data rates beyond $1$ Gigabit/s. Appropriate \MIMO\ detection schemes, offering a range of performance/complexity tradeoffs \cite{Larsson09} - and also joint iterative detection and decoding schemes \cite{Hagenauer2002}, have been essential for the technological improvements in this area. In the last decade, \MU-\MIMO\ systems have been successfully implemented and introduced in several broadband communication standards \cite{Gesbert07}; in such "space division mutiple access" systems, the more antennas the BS is equipped with, the more users can simultaneously communicate in the same time-frequency resource.

Recently, the adoption of \MU-\MIMO\ systems with a very large number of antennas in the BS, much larger than the number of mobile terminal (MT) antennas in its cell, was proposed in \cite{Marzetta10}. This "massive \MIMO" approach has been shown to be recommendable for several reasons \cite{Marzetta10,Rusek13,Hoydis13}: simple linear processing for \MIMO\ detection/precoding (uplink/downlink), namely when using \OFDM\ for broadband block transmission, becomes nearly optimal; both \MUI/\MSI\ effects and fast fading effects of multipath propagation tend to disappear; both power efficiency and bandwidth efficiency become substantially increased.

This paper deals with \SC/\FDE\ for the uplink of a \MU-\MIMO\ system where the BS is constrained to adopt low-complexity detection techniques but can be equipped with a large number of receiver antennas. In this context, either a linear detection or a reduced-complexity iterative \DF\ detection are considered. As to the linear detection alternative, we include both the optimum \MMSE\ \cite{Kim2008} and the quite simple \MF\ detection cases. The iterative \DF\ detection alternative, which resorts to joint cancellation of estimated \MUI/\MSI\ and \ISI, does not involve channel decoding,differently from the iterative receiver technique of \cite{Hagenauer2002}; it can be regarded as an extension to the multi-input context of the reduced-complexity iterative receiver techniques previously considered for \SIMO\ systems by the authors (see \cite{PT2006ISTC,GC2007_Revista2,GC2009_GLOBECOM} and the references therein).

Regarding performance evaluation by simulation, appropriate semi-analytical methods are proposed, combining simulated channel realizations and analytical computations of BER performance which are conditional on those channel realizations; selected analytical and semi-analytical performance bounds and a simple characterization of ''massive \MIMO'' effects are also provided. This paper shows and discusses a set of numerical performance results. The main conclusions of the paper are presented in the final section.

\section{System Model}\label{sec2}

\subsection{SC/FDE for MU-MIMO Uplink Block Transmission}\label{subsec2A}

We consider here a \CP-assisted \SC/\FDE\ block transmission, within a \MU-\MIMO\ system with $N_T$ TX antennas and $N_R$ RX antennas; for example, but not necessarily, one antenna per \MT. We assume, in the $j$th TX antenna ($j=1,2,...,N_T$) a length-$N$ block $s^{(j)}=[s_0^{(j)}, s_1^{(j)},...,s_{N-1}^{(j)}]^T$  of time-domain data symbols in accordance with the corresponding binary data block and the selected 4-QAM constellation under a Gray mapping rule. The insertion of a length-$L_s$ \CP, long enough to cope with the time-dispersive effects of multipath propagation, is also assumed.

By using the frequency-domain version of the time-domain data block $\bold{s}^{(j)}$, given by $\bold{S}^{(j)}=\left[S_0^{(j)},S_1^{(j)},\cdots,S_{N-1}^{(j)}\right]^T = DFT\left(\bold{s}^{(j)}\right)$ $(j=1,2,\cdots,N_T)$, we can describe the frequency-domain transmission rule as follows, for any subchannel $k$ $\left(k=0,1,\cdots, N-1\right)$:
\begin{eqnarray}\label{eq1}
   \bold{Y}_k = \bold{H}_k \bold{S}_k + \bold{N}_k,
\end{eqnarray}
where $\bold{S}_k = \left[S_k^{(1)},S_k^{(2)},\cdots,S_k^{(N_T)}\right]^T$ is the ''input vector'', $\bold{N}_k = \left[N_k^{(1)},N_k^{(2)},\cdots,N_k^{(N_R)}\right]^T$ is the Gaussian noise vector $\left(E\left[N_k^{(i)}\right]=0\right.$ and $\left.E\left[\left|N_k^{(i)}\right|^2\right]=\sigma_N^2=N_0 N\right)$, $\bold{H}_k$ denotes the $N_R\times N_T$ channel matrix with entries $H_k^{(i,j)}$, concerning a given channel realization, and $\bold{Y}_k= \left[Y_k^{(1)},Y_k^{(2)},\cdots,Y_k^{(N_R)}\right]^T$ is the resulting, frequency-domain, ''output vector'' .

As to a given \MIMO\ channel realization, it should be noted that the \CFR\ $\bold{H}^{(i,j)}=\left[H_0^{(i,j)}, H_1^{(i,j)},...,H_{N-1}^{(i,j)}\right]^T$, concerning the antenna pair $(i,j)$, is the DFT of the \CIR\ $\bold{h}^{(i,j)}=\left[h_0^{(i,j)}, h_1^{(i,j)},...,h_{N-1}^{(i,j)}\right]^T$, where $h_n^{(i,j)}=0$ for $n > Ls\ (n=0,1,...,N-1)$. Regarding a statistical channel model - which encompasses all possible channel realizations -, let us assume that $E\left[h_{n}^{(i,j)}\right]=0$ and $E\left[h_n^{(i,j)*}h_{n'}^{(i,j)}\right]=0$ for $n'\neq n$. By also assuming, for any $(i,j,k)$, a constant
\begin{eqnarray}\label{eq2}
   E\left[\left|H_k^{(i,j)}\right|^2\right] = \sum_{n=0}^{N-1} E\left[ \left| h_n^{(i,j)} \right|^2 \right] = P_\Sigma
\end{eqnarray}
(of course, with $h_n^{(i,j)}=0$ for $n>L_s$), the average bit energy at each \BS\ antenna is given by
\begin{eqnarray}\label{eq3}
   E_b = \frac{\sigma_s^2}{2\eta} P_\Sigma = \frac{\sigma_S^2}{2\eta N} P_\Sigma,
\end{eqnarray}
where $\eta=\frac{N}{N+L_s}$, $\sigma_S^2=E\left[\left|S_k^{(j)}\right|^2\right]$ and $\sigma_s^2=E\left[\left|s_n^{(j)}\right|^2\right] = \frac{\sigma_S^2}{N}$.

\subsection{Linear Detection Techniques}\label{subsec2B}

\begin{figure}[!ht]
\includegraphics[width=1\linewidth]{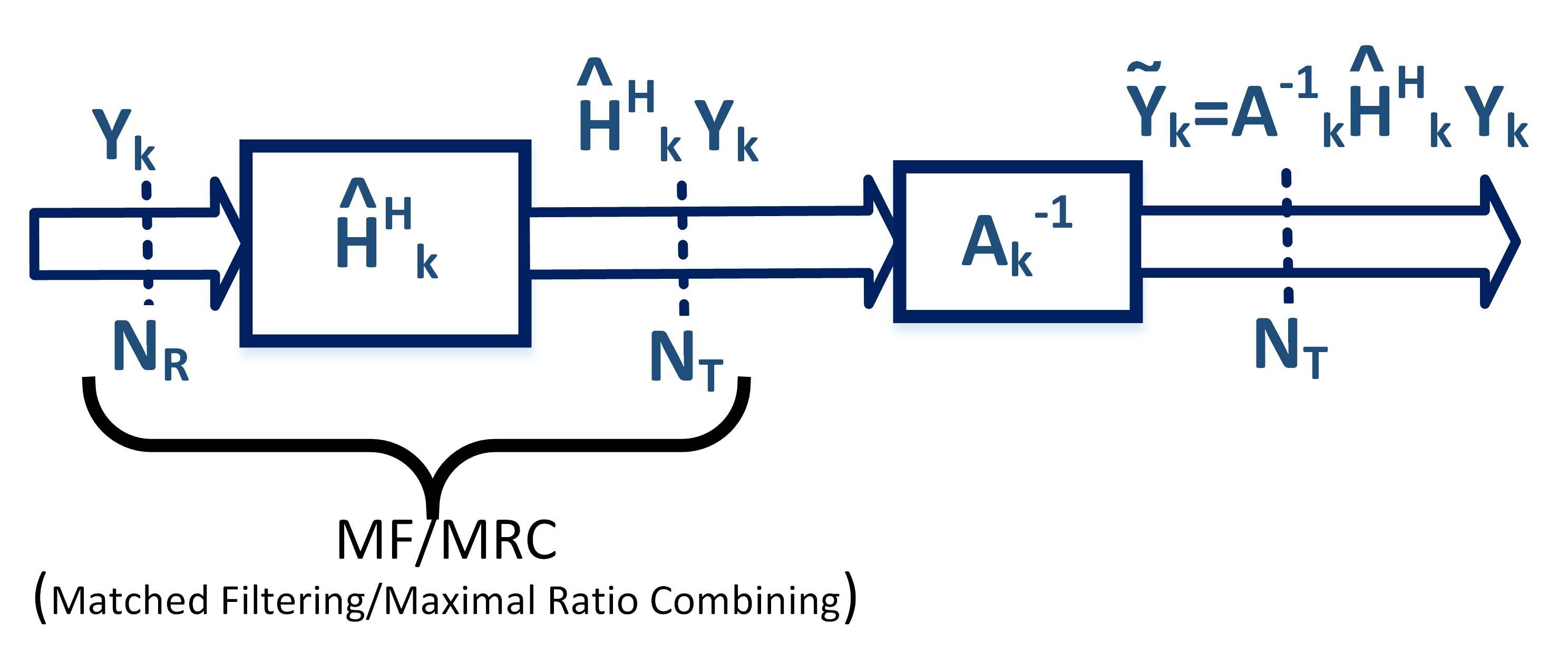} \vspace{-.50cm}
\\\centering{\bf (a)} \vspace{-0.0cm}\\
\includegraphics[width=1\linewidth]{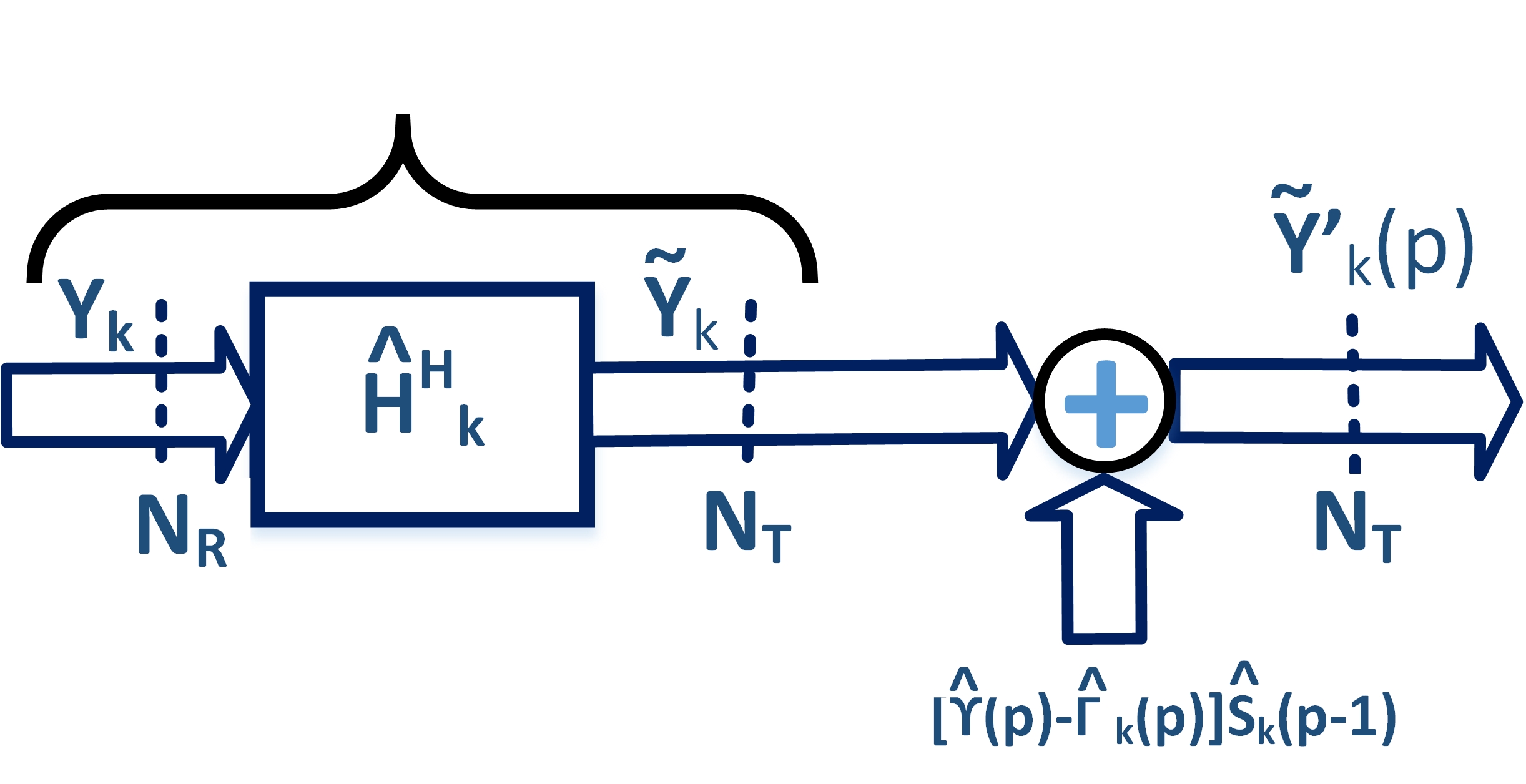} \vspace{-.60cm}
\\ \centering{\bf (b)} \vspace{-.25cm}\\
\caption{Frequency-domain linear detection (a) and iterative DF detection combining a linear MF detection and interference cancellation in the frequency domain (b).}   \label{OFDM_MIMO_Block2}
\end{figure}

An appropriate linear detector can be implemented by resorting to frequency-domain processing. After CP removal, a DFT operation leads to the required set $\left\{\bold{Y}_k; \ k=0,1,\cdots,N-1\right\}$ of length-$N_R$ inputs to the frequency-domain detector ($\bold{Y}_k$ given by (\ref{eq1})); it works, for each $k$, as shown in Fig. \ref{OFDM_MIMO_Block2}(a), leading to a set $\left\{\bold{\tilde{Y}}_k; \ k=0,1,\cdots,N-1\right\}$ of length-$N_T$ outputs $\bold{\tilde{Y}}_k=\left[\tilde{Y}_k^{(1)},\tilde{Y}_k^{(2)},\cdots,\tilde{Y}_k^{(N_T)}\right]^T \ (k=0,1,\cdots,N-1)$.

When $N_T \leq N_R$, possibly with $N_R \gg 1$, either an \MMSE, frequency-domain, optimum linear detection or a reduced-complexity,  frequency-domain, linear detection can be considered. In all cases, the detection matrix, for each subchannel $k$ ($k=0,1,...,N-1$) can be written as 
\begin{eqnarray}\label{eq4}
\textbf{D}_k =  \textbf{A}_k^{-1} \widehat{\textbf{H}}_k^H, 
\end{eqnarray} 
where $\widehat{\textbf{H}}_k^H$  is the conjugate transpose of the estimated \MU-\MIMO\ channel matrix $\widehat{\bold{H}}_k$ and $\bold{A}_k$ is a selected $N_T\times N_T$ matrix, possibly depending on $\widehat{\textbf{H}}_k$.  Therefore, $\bold{\tilde{Y}}_k = \bold{D}_k \bold{Y}_k = \bold{A}_k^{-1} \widehat{\bold{H}}_k^{H} \bold{Y}_k$
at the output of the frequency-domain linear detector (see Fig. \ref{OFDM_MIMO_Block2}(a)).

It should be noted that the $jth$ component of $\bold{\widehat{H}}_k^{H} \bold{Y}_k$ is given by $\sum\limits_{i=1}^{N_R} \widehat{H}_k^{(i,j)*} Y_k^{(i)}$ $(j=1,2,\cdots, N_T)$: this means that the $\widehat{\textbf{H}}_k^H$ factor provides $N_T$ \MRC\ procedures, one per \MT\ antenna, all of them based on an appropriate \MF\ for each component of the length-$N_R$ received vector at subchannel $k$.

For a \MMSE\ detection - the optimum linear detection - or a \ZF\ detection \cite{Larsson09,Kim2008}, an inversion of each $N_T\times N_T$ $\bold{A}_k$ matrix  $\left(k=0,1,\cdots, N-1\right)$ is required.

A reduced-complexity linear detection can be achieved by using an $N_T\times N_T$ diagonal matrix $\bold{A}_k$. The easiest implementation corresponds to adopting an identity matrix $\bold{A}_k = \bold{I}_{N_T}$. Of course, $\bold{D}_k = \widehat{\bold{H}}_k^H$ and $\widetilde{\bold{Y}}_k = \widehat{\bold{H}}_k^H \bold{Y}_k$ when  $\bold{A}_k = \bold{I}_{N_T}$, which means an ''\MF\ detection'', actually not requiring a matrix inversion.

For a given channel realization $\bold{H}_k$ and a given detection matrix $\bold{D}_k$, which depends on the estimated channel realization $\bold{\widehat{H}}_k$, the output of the frequency-domain detector is given by
\begin{eqnarray}\label{eq5}
 \bold{\tilde{Y}}_k = \bold{D}_k \bold{Y}_k = \bold{\Gamma}_k \bold{S}_k + \bold{N'}_k,
\end{eqnarray}
where  $\bold{\Gamma}_k = \bold{D}_k \bold{H}_k $ and $\bold{N'}_k = \bold{D}_k \bold{N}_k$.

With \SC/\FDE\ (time-domain data symbols), an \IDFT\ is required for each  $\bold{\widetilde{Y}}^{(j)} = \left[\widetilde{Y}_0^{(j)},\widetilde{Y}_1^{(j)},\cdots,\widetilde{Y}_{N-1}^{(j)}\right]^T$ vector. The $n$th component of the resulting length-$N$ $IDFT\left(\bold{\widetilde{Y}}^{(j)} \right) = \bold{\widetilde{y}}^{(j)}$ vector can be written as
\begin{eqnarray}\label{eq6}
    \widetilde{y}_n^{(j)} = \gamma^{(j)}s_n^{(j)} + ISI + MUI + \ 'Gaussian \ noise',
\end{eqnarray}
with $\gamma^{(j)} = \frac{1}{N}\sum\limits_{k=0}^{N-1} \Gamma_k^{(j,j)}$ \hspace{1cm} $\left( \gamma^{(j)} = \frac{E\left[ \widetilde{y}_n^{(j)} s_n^{(j)*}\right] }{\sigma_s^2}\right) $.

Therefore, $\bold{\widetilde{Y}}_k$ can be written as 
\begin{eqnarray}\label{eq7}
    \bold{\widetilde{Y}}_k  = \boldsymbol{\gamma} \bold{S}_k + \left(\bold{\Gamma}_k - \boldsymbol{\gamma}\right)\bold{S}_k + \bold{D}_k \bold{N}_k,
\end{eqnarray}
where $\boldsymbol{\gamma}$ is a diagonal $N_T\times N_T$ matrix with $(j,j)$ entries given by $\gamma^{(j)} = \frac{1}{N} \sum\limits_{k=0}^{N-1} \Gamma_k^{(j,j)}$.

When $\widetilde{Y}_k^{(j)}$ is written as
\begin{eqnarray}\label{eq8}
    \tilde{Y}_k^{(j)} = \gamma^{(j)} S_k^{(j)} + \left[\Gamma_k^{(j,j)} - \gamma^{(j)}\right]S_k^{(j)} + \\ \nonumber
    \begin{array}{c} \sum\limits_{l=1}^{N_T}\\*[-0.1cm]_{(l\neq j)}  \end{array} \Gamma_k^{(j,l)} S_k^{(l)} + \sum\limits_{i=1}^{N_R} D_k^{(j,i)} N_k^{(i)},
\end{eqnarray}
the four terms in the right-hand side of eq. (\ref{eq8}) are concerned, respectively, to ''signal'', \ISI, \MUI/\MSI\ and ''Gaussian noise'', at subchannel $k$.

\subsection{Low-Complexity Iterative DF Technique}\label{subsec2C}

A low-complexity iterative \DF\ technique can be easily devised having in mind eq.  (\ref{eq7}). This frequency-domain nonlinear detection technique combines the use of a linear detector and, for all iterations after the initial iteration (i.e., for $p>1$), a cancellation of residual \MUI\ - and residual \MSI, when some users adopt several TX antennas for spatial multiplexing purposes - as well as residual \ISI; such cancellation is based on the estimated data block which is provided by the preceding iteration and fed back to the frequency-domain detector. The output of this frequency-domain detector, for iteration $p$, is as follows:
\begin{eqnarray}\label{eq9}
    \bold{\widetilde{Y}}'_k (p) = \bold{\widetilde{Y}}_k + \left[ \boldsymbol{\widehat{\gamma}}(p) - \bold{\widehat{\Gamma}}_k(p)\right] \bold{\widehat{S}}_k (p-1),
\end{eqnarray}
$\left[k=0,1,\cdots, N-1; p>1\right.$ (for $\left. \left. p=1,  \bold{\widetilde{Y}}'_k (p) = \bold{\widetilde{Y}}_k\right) \right]$, where $\bold{\widehat{\Gamma}}_k (p) = \bold{D}_k (p) \bold{\widehat{H}}_k$ - with $\bold{D}_k (p)$ denoting the detection matrix employed in iteration $p$ - and the entries $(j,j)$ of the diagonal matrix $\widehat{\boldsymbol{\gamma}} (p)$ are given by $\widehat{\gamma}^{(j)} (p) = \frac{1}{N} \sum\limits_{k=0}^{N-1} \widehat{\Gamma}_k^{(j,j)} (p)$. Of course, $\left[\widehat{S}_0^{(j)}(p-1),\cdots,\widehat{S}_{N-1}^{(j)}(p-1)\right]^T = DFT\left(\left[\widehat{s}_0^{(j)}(p-1),\cdots,\widehat{s}_{N-1}^{(j)}(p-1)\right]^T\right)$.

The implementation of this iterative \DF\ technique is especially simple when $\bold{D}_k (p) = \bold{\widehat{H}}_k^H$ for any $p$, i.e., when a linear \MF\ detector is adopted as shown in Fig. \ref{OFDM_MIMO_Block2} (b) for all iterations; the matrix inversion which is inherent to more sophisticated linear detectors is then avoided. On the other hand, a slightly improved performance can be achieved through an increased  implementation complexity, by feeding back vectors of soft (instead of hard) time-domain symbol decisions for interference cancellation.

\section{Evaluation of the Achievable Detection Performances}\label{sec3}

\subsection{Semi-analytical Performance Evaluation}\label{subsec3A}

Regarding evaluation of detection performances by simulation, simple semi-analytical methods are presented here, for the detection techniques of subsecs \ref{subsec2B} and \ref{subsec2C}, both combining simulated channel realizations and analytical computations of \BER\ performance which are conditional on those channel realizations. In all cases, the conditional \BER\ values are directly computed by resorting to a \SINR, under the assumption that the ''interference'' has a quasi-Gaussian nature. These ratios are simply derived in accordance with the channel realization $\bold{H}_k$ ($k=0,1,\cdots, N-1$). Of course, for concluding the \BER\ computation in each case - involving random generation of a large number of channel realizations and conditional \BER\ computations - a complementary averaging operation over the set of channel realizations is required.

When using a linear detection technique (sec. \ref{subsec2B}), it is easy to conclude, having in mind  (\ref{eq8}), that the ''signal-to-interference-plus-noise'' ratio concerning the $j$th input of the MU-MIMO system is given by

%
\begin{eqnarray}\label{eq10}
 SINR_{j} =  
&\frac{N\left|\bold{\gamma}^{(j)}\right|^2}{\beta_j + \begin{array}{c} \sum\limits_{l=1}^{N_T}\\*[-0.1cm]_{(l\neq j)}  \end{array}\beta_l  + \alpha\sum\limits_{i=1}^{N_R}\sum\limits_{k=0}^{N-1}\left|D_k^{(j,i)}\right|^2}
\end{eqnarray}
where $\alpha = \frac{N_0}{\sigma_s^2}$, $\beta_j = \sum\limits_{k=0}^{N-1}\left|\Gamma_k^{(j,j)}-\gamma^{(j)}\right|^2$ and  $\beta_l = \sum\limits_{k=0}^{N-1}\left|\Gamma_k^{(j,l)}\right|^2$ with $l\neq j$.

For 4-QAM transmission, the resulting $BER_j$ ($j=1,2,\cdots, N_T$) - conditional on the channel realization $\{\bold{H}_k; \ k = 0,1,\cdots,N-1\}$ - is given by 
\begin{eqnarray}\label{eq11}
    BER_j \approx Q\left(\sqrt{SINR_j}\right),
\end{eqnarray}
(where $Q(.)$ is the Gaussian error function) with $SINR_j$ as computed above, and $BER = \frac{1}{N_T}\sum\limits_{j=1}^{N_T} BER_{j}$.

When using the iterative \DF\ technique of Fig. \ref{OFDM_MIMO_Block2} (b), eq. (\ref{eq11}) is replaced by $BER_j (p) \approx Q\left(\sqrt{SINR_j(p)}\right)$ for $p\geq 1$; certainly, $\bold{D}_k (p) = \widehat{\bold{H}}_k^H$ and $\bold{\Gamma}_k (p) = \widehat{\bold{H}}_k^H \bold{H}_k$ in the computation of $SINR_j (p)$. By assuming a perfect channel estimation $\left(\widehat{\bold{H}}_k = \bold{H}_k\right)$, this computation for $p>1$ can be made especially simple, taking into account that $\widetilde{\bold{Y}}' (p) = \boldsymbol{\gamma} \bold{S}_k + \left( \boldsymbol{\Gamma}_k - \boldsymbol{\gamma}\right) \left(\bold{S}_k - \widehat{\bold{S}}_k (p-1)\right) + \bold{D}_k \bold{N}_k$ and $E\left(\left|s_n^{(j)} - \widehat{s}_n^{(j)} (p-1)\right|^2\right) = 4 \sigma_s^2 BER_j (p-1) \approx 4\sigma_s^2 Q\left(\sqrt{SINR_j (p-1)}\right)$. Therefore, we get
\begin{eqnarray}\label{eq12}
 SINR_{j} (p) \approx  
&\frac{N\left|\bold{\gamma}^{(j)}\right|^2}{\beta_j + \begin{array}{c} \sum\limits_{l=1}^{N_T}\\*[-0.1cm]_{(l\neq j)}  \end{array}\beta_l  + \alpha\sum\limits_{i=1}^{N_R}\sum\limits_{k=0}^{N-1}\left|D_k^{(j,i)}\right|^2}
\end{eqnarray}
where $\beta_l = 4Q\left(\sqrt{SINR_l (p-1)}\right)\sum\limits_{k=0}^{N-1}\left|\Gamma_k^{(j,l)}\right|^2$ with $l\neq j$ and $\beta_j = 4Q\left(\sqrt{SINR_j (p-1)}\right)\sum\limits_{k=0}^{N-1}\left|\Gamma_k^{(j,j)}-\gamma^{(j)}\right|^2$.

\subsection{Reference MMSE Performance and SIMO Performance Bounds}\label{subsec3B}

When adopting an ''\MMSE\ detector'' and a perfect channel estimation is assumed, $\bold{D}_k = \left( \bold{H}_k^H \bold{H}_k + \alpha I_{N_T}\right)^{-1} \bold{H}_k^H$ \cite{Kim2008}. It can be shown that the resulting $SINR_j$ - which can be used for computing $BER_j$, and then \BER\ - can be written as $SINR_j = \frac{\gamma^{(j)}}{1-\gamma^{(j)}}$, with $\gamma^{(j)}$ as defined in subsec.  \ref{subsec2B}.

\begin{table}[!ht]
\caption{SIMO ($1 \times N_R$) Performance Bounds.}
\centering\begin{tabular}[]{|l||l|}
\hline
SIMO/LDB &  $SINR_{1} = \frac{1}{\alpha}\frac{ \sum\limits_{k=0}^{N-1} \frac{\sum\limits_{i=1}^{N_R}\left|H_k^{(i,1)}\right|^2}{\alpha + \sum\limits_{i=1}^{N_R}\left|H_k^{(i,1)}\right|^2}}{ \sum\limits_{k=0}^{N-1} \frac{1}{\alpha + \sum\limits_{i=1}^{N_R}\left|H_k^{(i,1)}\right|^2}}$ \\ \hline
SIMO/MFB  &  $SINR_{1} = \frac{1}{\alpha N}\sum\limits_{k=0}^{N-1}\sum\limits_{i=1}^{N_R} \left|H_k^{(i,1)}\right|^2$  \\ \hline
SIMO/AWGN/MFB  &$ SINR_{1} = 2\eta N_R \frac{E_b}{N_0}$ \\ \hline
\end{tabular}\label{table1}
\end{table}

Successively improved performance bounds can be obtained as follows (see Table \ref{table1}): also under  $\bold{D}_k = \left(\bold{H}_k^H \bold{H}_k + \alpha \bold{I}_{N_T}\right)^{-1} \bold{H}_k^H$, with the same $N_R$ but $N_T=1$, which corresponds to a \SIMO/\LDB; under $\bold{D}_k = \bold{H}_k^H$, with $N_T=1$ and the same $N_R$, by suppressing the resulting first term (\ISI) in the denominator of eq. (\ref{eq10}), which corresponds to a \SIMO/\MFB; when replacing the fading channel by an AWGN channel for each $(i,1)$ antenna pair $(i=1,2,\cdots, N_R)$ in the \SIMO/\MFB, which corresponds to a \SIMO/AWGN/\MFB.

It should be noted that the BER curve for the \SIMO/AWGN/\MFB\ actually corresponds to the achievable BER performance (against White Gaussian Noise) in a \SIMO\ $1\times N_R$ system with single-path propagation for all $(i,1)$ antenna pairs, provided that an \MF\ detection, under ideal channel estimation, is adopted:
\begin{eqnarray}\label{eq13}
    BER = Q \left(\sqrt{2\eta N_R \frac{E_b}{N_0}}\right)
\end{eqnarray}
%

\begin{figure*}[!ht]
\begin{minipage}[t]{0.325\linewidth}
\centering\includegraphics[width=.8\linewidth]{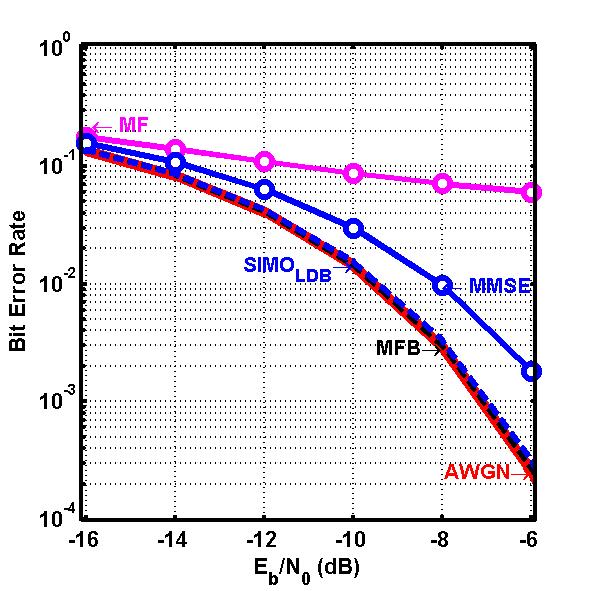} \vspace{-0.125cm}
\\ \centering{\bf(a)}
\end{minipage} \hfill\vspace{-.00cm}
\begin{minipage}[t]{0.325\linewidth}
\centering\includegraphics[width=.8\linewidth]{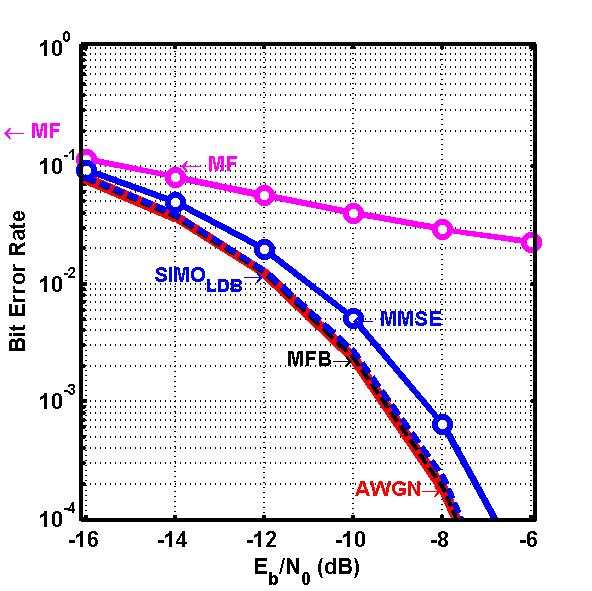} \vspace{-0.125cm}
\\\centering{\bf(b)}
\end{minipage}\hfill\vspace{-.00cm}
\begin{minipage}[t]{0.325\linewidth}
\centering\includegraphics[width=.8\linewidth]{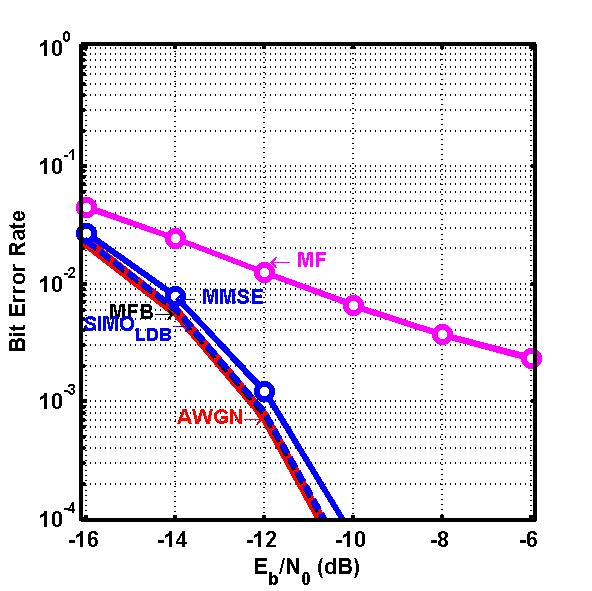} \vspace{-0.125cm}
\\\centering{\bf(c)}
\end{minipage}\hfill\vspace{-.025cm}
\caption{BER performances for \SC/\FDE-based \MU-\MIMO, with $N_T=10$ and $N_R=30$ (a), $50$ (b) or  $100$  (c), when using linear detection (\MF\ and \MMSE) [with the \SIMO/\LDB\ (dashed line), the \SIMO/\MFB\ and the \SIMO/AWGN/\MFB\   $(1\times N_R)$ performances also included].}  \label{Fig:simsSC_met5_Nt10}
\end{figure*}

\begin{figure*}[!ht]
\begin{minipage}[t]{0.325\linewidth}
\centering\includegraphics[width=.8\linewidth]{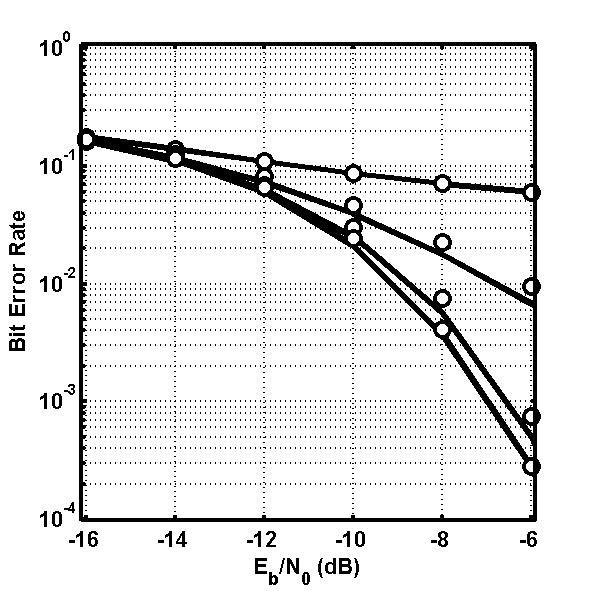} \vspace{-0.125cm}
\\ \centering{\bf(a)}
\end{minipage} \hfill\vspace{-.00cm}
\begin{minipage}[t]{0.325\linewidth}
\centering\includegraphics[width=.8\linewidth]{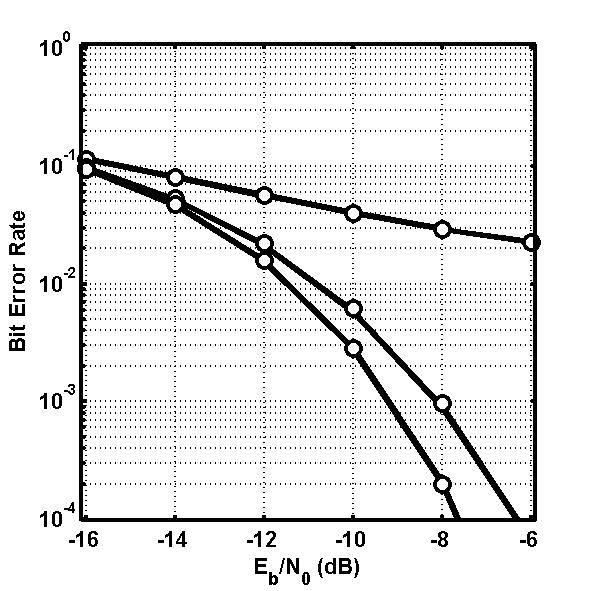} \vspace{-0.125cm}
\\\centering{\bf(b)}
\end{minipage}\hfill\vspace{-.00cm}
\begin{minipage}[t]{0.325\linewidth}
\centering\includegraphics[width=.8\linewidth]{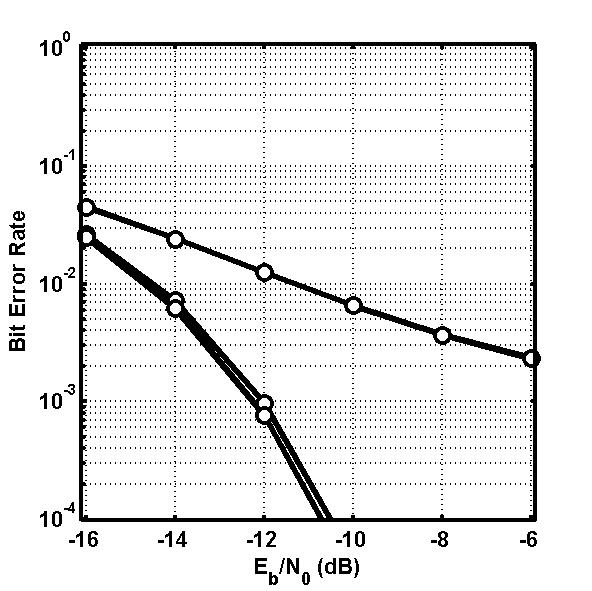} \vspace{-0.125cm}
\\\centering{\bf(c)}
\end{minipage}\hfill\vspace{-.025cm}
\caption{BER performances, at the several iterations ($p=1,2,3,4$) for \SC/\FDE-based \MU-\MIMO, with $N_T=10$ and $N_R=30$ (a), $50$ (b) or  $100$  (c), when using the iterative DF detection technique.}  \label{Fig:simsSC_hard_Nt10}
\end{figure*}

\subsection{Massive MIMO effects}\label{subsec3C}

When $N_R \gg N_T$, both the \MUI/\MSI\ effects and the effects of multipath propagation (fading, \ISI) tend to disappear: consequently, the \BER\ performances for the \MU-\MIMO\ $N_T \times N_R$ Rayleigh fading channel become very close  to those concerning a \SIMO\  $1\times N_R$ channel with single-path propagation for all $N_R$ TX/RX antenna pairs.  The achievable performances under a "truly massive" \MU-\MIMO\ implementation can be analytically derived as explained in the following. 

Entries of $\bold{H}_k$ are i.i.d. Gaussian-distributed random variables with zero mean and variance $P_\Sigma$. Therefore, $\lim\limits_{N_R\rightarrow\infty} \left[\frac{1}{N_R} \sum\limits_{i=1}^{N_R}\left|H_k^{(i,j)}\right|^2\right] = E\left[\left|H_k^{(i,j)}\right|^2\right] = P_\Sigma$ and
\begin{eqnarray}\label{eq13a}
    &&\lim\limits_{N_R\rightarrow\infty} \left[\frac{1}{N_R} \begin{array}{c}\sum\limits_{i=1}^{N_R} \\*[-0.1cm]_{(l\neq j)} \end{array} H_k^{(i,j)*} H_k^{(i,l)}\right] \\
    &&=  \begin{array}{c} E \\*[-0.2cm]_{l\neq j}\end{array} \left[H_k^{(i,j)*} H_k^{(i,l)}\right] = 0. \nonumber
\end{eqnarray}

Consequently, for $N_R\gg N_T$, 
\begin{eqnarray}\label{eq14}
    \sum\limits_{i=1}^{N_R} \left|H_k^{(i,j)}\right|^2 \approx N_R P_\Sigma
\end{eqnarray}
and
\begin{eqnarray}\label{eq15}
    \begin{array}{c}\sum\limits_{i=1}^{N_R} \\*[-0.1cm]_{(l\neq j)} \end{array} H_k^{(i,j)*} H_k^{(i,l)} \approx 0.
\end{eqnarray}

Therefore,
\begin{eqnarray}\label{eq16}
     \begin{array}{c} lim \\*[-0.2cm]_{N_R\rightarrow \infty}\end{array}\left(\frac{SINR_{j}}{N_R}\right) = \frac{\sigma_s^2}{N_0}P_\Sigma = 2\eta \frac{E_b}{N_0}
\end{eqnarray}
(by assuming that $\bold{D}_k = \bold{H}_k^H$). When $N_R \gg N_T$, 
\begin{eqnarray}\label{eq17}
     SINR_j & \approx N_R \begin{array}{c} lim \\*[-0.2cm]_{N_R\rightarrow \infty}\end{array}\left(\frac{SINR_{j}}{N_R}\right) \\ \nonumber
     & = 2\eta N_R \frac{E_b}{N_0},
\end{eqnarray}
which implies that $BER \approx Q\left(\sqrt{2\eta N_R \frac{E_b}{N_0}}\right)$, i.e. a BER performance closely approximating the \SIMO/AWGN/MFB (eq. (\ref{eq13})).

When $N_R\gg N_T$, it should also be noted - having in mind the equations (\ref{eq14}) and (\ref{eq15}) - that the \MMSE\ linear detection becomes practically equivalent to a \MF\ linear detection, since 
\begin{eqnarray}\label{eq18}
    \bold{D}_k = \left(\bold{H}_k^H \bold{H}_k + \alpha \bold{I}_{N_T}\right)^{-1} \bold{H}_k^H \approx \beta \bold{H}_k^H,
\end{eqnarray}
with $\beta = \frac{1}{\alpha + N_R P_\Sigma}$ (assuming a perfect channel estimation). Of course, the corresponding $\bold{\Gamma}_k$ matrix is then $\bold{\Gamma}_k\approx \beta \bold{H}_k^H  \bold{H}_k \approx \frac{N_R P_\Sigma}{\alpha + N_R P_\Sigma} \bold{I}_{N_T}$, leading to $\gamma^{(j)}\approx \Gamma_k^{(j,j)} \approx \frac{N_R P_\Sigma}{\alpha + N_R P_\Sigma}$ ($j=1,2,\cdots, N_T$). Therefore, the resulting \SINR 's are as expected, under a perfect channel estimation $\left(SINR_{j} = \frac{\gamma^{(j)}}{1-\gamma^{(j)}} \approx \frac{N_R P_\Sigma}{\alpha}= 2\eta N_R \frac{E_b}{N_0}\right)$. Clearly, when $N_R\gg N_T$, the \MUI/\MSI\ effects as well as both the fading and the ISI effects of multipath propagation become vanishingly small, leading to a close approximation to the \SIMO/AWGN/\MFB\ reference performance.


\section{Numerical Results and Discussion}\label{sec4}

The set of performance results which are presented here are concerned to \SC/\FDE\ uplink block transmission, with $N=256$ and $Ls=64$ in a \MU-\MIMO\ $N_T\times N_R$ Rayleigh fading channel. Perfect channel estimation and perfect power control are assumed. The fading effects regarding the several TX/RX antenna pairs are supposed to be uncorrelated, with independent zero-mean complex Gaussian $h_n^{(i,j)}$ coefficients assumed to have variances $P_n = 1 - \frac{n}{63}$, $n=0,1,...,63$ ($P_n=0$ for $n=64,65,...,255$).

The accuracy of performance results obtained through the semi-analytical simulation methods of sec. \ref{sec3} was assessed by means of parallel conventional Monte Carlo simulations (involving an error counting procedure), which correspond to the superposed dots in the several \BER\ performance curves of Figs. \ref{Fig:simsSC_met5_Nt10} and \ref{Fig:simsSC_hard_Nt10}, concerning the \MU-\MIMO\ system.

Fig. \ref{Fig:simsSC_met5_Nt10} shows the simulated \BER\ performances for an \SC/\FDE-based \MU-\MIMO\ uplink and three possibilities regarding $N_R$ for $N_T=10$, when using two linear detection techniques: optimum (\MMSE) detection; reduced-complexity (\MF) detection.  In each subfigure, for the sake of comparisons, we also include the \SIMO\ $1 \times N_R$ performance bounds of Table \ref{table1}. In the simulation results concerning each subfigure of  Fig. \ref{Fig:simsSC_met5_Nt10}, the five BER performance curves are ordered, from the worst to the best, as follows: $N_T \times N_R$  \MU-\MIMO\ with reduced-complexity (\MF) linear detection; $N_T \times N_R$ \MU-\MIMO\ with \MMSE\ detection;  \SIMO/\LDB\ ($1\times N_R$); \SIMO/\MFB\ ($1\times N_R$); \SIMO/AWGN/\MFB\ ($1\times N_R$) [practically superposed to the \SIMO/\MFB\ curve]. These results clearly show that the performance degradation which is inherent to the reduced-complexity linear detection technique (\MF) - as compared with the \MMSE\ linear detection - can be made quite small, by increasing $N_R$ significantly; they also show that, under highly increased $N_R$ values, the ''\MUI/\MSI-free'' \SIMO\ (multipath) performance and the ultimate bound - the ''\MUI/\MSI-free and \ISI\ $\&$ fading-free" \SIMO\ (single-path) performance - can be closely approximated, even when adopting the reduced-complexity linear detection.  This figure emphasizes a "massive \MIMO" effect when $N_R \gg  N_T$, which leads to \BER\ performances very close to the ultimate "\MUI/\MSI-free and \ISI\ $\&$ fading-free" \SIMO\ (single-path) performance bound (the \SIMO/AWGN/\MFB\ of Table \ref{table1}). 

Fig. \ref{Fig:simsSC_hard_Nt10} shows the simulated BER performances for an \SC/\FDE-based \MU-\MIMO\ uplink and three possibilities regarding $N_R$ for $N_T=10$, when using the reduced-complexity iterative \DF\ detection technique of Fig. \ref{OFDM_MIMO_Block2} (b), which does not require matrix inversions. By comparing these results to the results of Fig. \ref{Fig:simsSC_met5_Nt10}, also for $N_T=10$, we can conclude that, whenever $N_R\geq 3 N_T$, the reduced-complexity \DF\ technique of Fig. \ref{OFDM_MIMO_Block2} (b) is able to provide BER performances which are better than those of the \MMSE\ linear detector [while avoiding the inversion of a $10\times 10$ matrix $A_k$ $(k=0,1,\dots, N-1)$], closely approximating the (practically identical) \SIMO/\MFB\ and the \SIMO/AWGN/\MFB\ reference performances after a small number of iterations.

\begin{figure}[!ht]
\centering\includegraphics[width=1\linewidth]{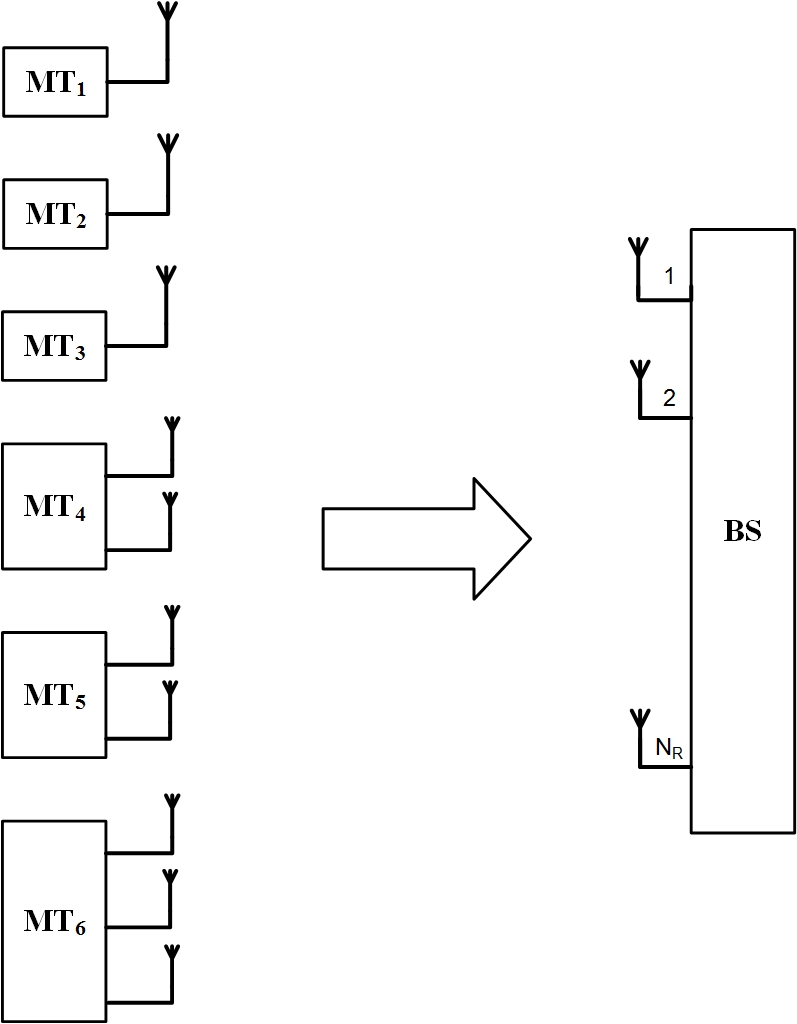} \vspace{-.1250cm}
\caption{Uplink transmission scenario with six users and $N_T=10$ TX antennas.}   \label{MIMO_TX_RX1}
\end{figure}

It should be noted that the numerical results reported above, for $N_T=10$ TX antennas, are compatible with an uplink transmission scenario involving up to $10$ users (for example, $10$ users with one TX antenna per user); a specific scenario with $N_T=10$ TX antennas, involving six users, is depicted in Fig. \ref{MIMO_TX_RX1}.

\section{Conclusions}\label{sec5}

This paper was dedicated to the uplink detection and performance evaluation for a \MU-\MIMO\ system with \SC/\FDE\ transmission, when adopting a large number of antennas and low-complexity detection techniques at the BS. With the help of selected numerical performance results, discussed in detail in Section \ref{sec4}, we show that a moderately large number of \BS\ antennas (say, $N_R = 3 N_T$) is enough to closely approximate the \SIMO/\MFB\ performance - and also the \SIMO/AWGN/\MFB\ performance, expressed as $BER = Q\left(\sqrt{2\eta N_R \frac{E_b}{N_0}}\right)$ -, especially when using the suggested low-complexity iterative \DF\ technique, which does not require $N_T\times N_T$ matrix inversion. We also emphasize the "massive \MIMO" effects provided by a number of BS antennas much higher than the number of antennas which are jointly employed in the terminals of the multiple autonomous users, even when strongly reduced-complexity linear detection techniques - such as the so-called ''\MF\ detection -, are adopted.

The accuracy of performance results obtained by semi-analytical means, much less time-consuming than conventional, 'error counting'-based, Monte Carlo simulations - was also demonstrated. The proposed performance evaluation method can be very useful for rapidly knowing "how many antennas do we need in the \BS?", for a given number of antennas jointly employed in the user terminals.

\bibliographystyle{unsrt}    
\bibliography{LIV_massive} 

\end{document}